\documentclass[preprint,aps,pre,superscriptaddress,nofootinbib]{revtex4-2}

\usepackage{amsmath,amssymb,amsfonts,bm,mathrsfs}
\usepackage{graphicx}
\usepackage{hyperref}
\usepackage{placeins}
\usepackage{caption}
\usepackage{subcaption}
\usepackage{mathtools}
\mathtoolsset{showonlyrefs}

\DeclareMathOperator{\sgn}{sgn}
\newcommand{\He}{\mathrm{He}}
\newcommand{\sqbra}[1]{\left[ #1 \right]}

\begin{document}

\title{Testing the Role of Diagonal Interactions in
High-Order Hopfield Models via Dynamical Mean-Field Theory}
\author{Yuto Sumikawa}
\email{sumikawa.yuto.p1@alumni.tohoku.ac.jp}
\affiliation{Department of Physics, The University of Tokyo, 7-3-1 Hongo, Bunkyo-ku, Tokyo 113-0033, Japan}
\author{Yoshiyuki Kabashima}
\email{kaba@phys.s.u-tokyo.ac.jp}
\affiliation{Institute for Physics of Intelligence \& Trans-Scale Quantum Science Institute, The University of Tokyo, 7-3-1 Hongo, Bunkyo-ku, Tokyo 113-0033, Japan}

\begin{abstract}
High-order extensions of the Hopfield model are known to exhibit dramatically enhanced storage capacity at equilibrium, while their dynamical retrieval properties remain less well understood.
In our previous work, we carried out a dynamical mean-field theory (DMFT) analysis of the Krotov--Hopfield-type dense associative memory and found that the transition between successful and failed retrieval is accompanied by pronounced slow dynamics. As a consequence, the effective basin of attraction observed in numerical simulations extends well beyond that predicted by equilibrium statistical mechanics.
A natural hypothesis is that this discrepancy originates from diagonal (self-interaction) contributions in the Krotov--Hopfield model, which generate a large number of lower-order interaction terms and may induce glassy relaxation near the retrieval boundary.
To test this hypothesis, we analyze an alternative high-order associative memory model, namely the Abbott--Arian-type $p$-body Hopfield model, in which such diagonal contributions are absent by construction.
Using dynamical mean-field theory, we derive an effective single-site process together with closed macroscopic equations governing the retrieval dynamics.
Our analysis reveals that both slow dynamics and a substantial enlargement of the apparent basin of attraction persist even in this model.
These results indicate that the dynamical slowdown near the retrieval boundary cannot be attributed primarily to diagonal self-interaction effects, but instead originates from intrinsic properties of high-order interactions.
\end{abstract}

\maketitle

\section{Introduction}

The Hopfield model \cite{Hopfield1982} is a paradigmatic associative memory model that has been extensively studied in statistical physics and theoretical neuroscience.
While the original model features pairwise interactions, extensions to higher-order interactions ($p$-body couplings) have been proposed to enhance storage capacity~\cite{Gardner1987,Abbott1987}.
The equilibrium properties of such models have been analyzed using replica theory, revealing a dramatic increase in storage capacity with $p$.
However, the dynamical aspects of retrieval, including convergence properties and basins of attraction, have received comparatively less attention beyond the $p=2$ case \cite{AmariMaginu1988,CoolenSherrington1993,Okada1995}.

In our previous work, we performed a dynamical mean-field theory (DMFT) analysis~\cite{Mimura2025} of the 
Krotov--Hopfield-type dense associative memory model~\cite{Krotov2016}.
We found that the transition between successful and failed retrieval is accompanied by pronounced slow dynamics.
As a consequence, the apparent basin of attraction observed in long but finite time numerical simulations extends well beyond that predicted by equilibrium statistical mechanics.
These findings indicate a significant discrepancy between static and dynamical characterizations of memory performance in high-order associative memory models.

A natural hypothesis is that this discrepancy originates from diagonal (self-interaction) contributions present in the 
Krotov--Hopfield model.
Such terms generate a large number of lower-order effective interactions and may induce glassy relaxation near the retrieval boundary.
It is therefore important to clarify whether these diagonal contributions are the primary source of the observed slow dynamics.

To address this question, in this paper we analyze an alternative high-order associative memory model, namely the Abbott--Arian-type $p$-body Hopfield model~\cite{Abbott1987}, in which diagonal contributions are absent by construction.
Although this model is defined through products of $p$ distinct Ising variables and thus appears structurally more involved, it provides a suitable framework for isolating the effect of diagonal terms.
Our analysis reveals that slow retrieval dynamics and a substantial enlargement of the apparent basin of attraction persist even in the absence of diagonal self-interaction terms.
This demonstrates that the dynamical slowdown near the retrieval boundary cannot be attributed primarily to such diagonal contributions, but instead originates from intrinsic properties of high-order interactions.

The remainder of this paper is organized as follows.
Section 2 introduces the model and notation.
Section 3 reviews the equilibrium storage-capacity result as a baseline.
Section 4 presents the DMFT derivation and the effective single-site process.
Section 5 reports numerical results for overlap dynamics and basins of attraction.
Section 6 concludes with implications and limitations, and Appendix A summarizes the technical details of the derivation.

\section{Model}

We consider a $p$-body extension of the Hopfield model with $N$ binary spins $\sigma_i(t)\in \{-1,+1\}$, updated synchronously at zero temperature.
The $p$-body Hebbian couplings are defined as
\begin{align}
J_{j_1,\dots,j_p}
= \frac{1}{N^{p-1}} \sum_{\mu=0}^{M-1}
\xi_{j_1}^{\mu} \xi_{j_2}^{\mu} \cdots \xi_{j_p}^{\mu},
\label{eq:J}
\end{align}
where the patterns ${\xi_i^\mu}$ are independent Rademacher random variables.

The dynamics is given by
\begin{align}
\sigma_i(t+1) = \operatorname{sgn}\bigl(h_i(t)\bigr),
\label{eq:update}
\end{align}
with the local field
\begin{align}
h_i(t)
= \sum_{j_2<\cdots<j_p}
J_{i,j_2,\dots,j_p}
\sigma_{j_2}(t)\cdots\sigma_{j_p}(t).
\label{eq:field}
\end{align}
The corresponding energy function (with distinct indices) is
\begin{align}
H(\boldsymbol{\sigma})
= - \sum_{j_1<\cdots<j_p}
J_{j_1,\dots,j_p},
\sigma_{j_1}\cdots\sigma_{j_p}.
\end{align}

We focus on the retrieval of pattern $\mu=0$ and define the overlap
\begin{align}
m(t) = \frac{1}{N} \sum_{i=1}^N \xi_i^0 \sigma_i(t).
\end{align}
Throughout the paper, we use the loading rate
\begin{align}
\alpha = \frac{M}{N^{p-1}}.
\end{align}

On the other hand, the Krotov--Hopfield model \cite{Krotov2016} is defined by the Hamiltonian
\begin{align}
H^{\mathrm{KH}}(\boldsymbol{\sigma})
= -\sum_{\mu=1}^M \frac{1}{2N^{p-1}}
\left( \sum_{i=1}^N \xi_i^\mu \sigma_i \right)^p.
\end{align}
The corresponding zero-temperature dynamics can be written as
\begin{align}
\sigma_i(t+1)
= \operatorname{sgn}\Biggl[\sum_{\mu=1}^M 
\left (
\frac{1}{2N^{p-1}}
\left( \sum_{j\neq i} \xi_j^\mu \sigma_j(t) + \xi_i^\mu \right)^p
- \frac{1}{2N^{p-1}}
\left( \sum_{j\neq i} \xi_j^\mu \sigma_j(t) - \xi_i^\mu \right)^p \right )
\Biggr].
\end{align}
Expanding this expression, one obtains
\begin{align}
\sigma_i(t+1)
= \operatorname{sgn}\left[
\sum_{\mu=1}^M \xi_i^\mu
\left( \frac{1}{N} \sum_{j\neq i} \xi_j^\mu \sigma_j(t) \right)^{p-1}
+ \mathcal{O}(N^{-1/2})
\right],
\label{eq:Krotov_dynamics}
\end{align}
where $\mathcal{O}(N^{-1/2})$ denotes subleading finite-size corrections.

To make the structure of interactions more explicit, we expand the $p$-th power in the Hamiltonian:
\begin{align}
\left( \sum_{i=1}^N \xi_i^\mu \sigma_i \right)^p
= \sum_{i_1,\dots,i_p=1}^N
\xi_{i_1}^\mu \cdots \xi_{i_p}^\mu \,
\sigma_{i_1} \cdots \sigma_{i_p}.
\end{align}
This sum includes not only terms with all indices distinct, but also terms in which some indices coincide.  
In particular, contributions with repeated indices generate effective lower-order interactions.  
For example, terms with one repeated index ($i_1 = i_2 \neq i_3,\dots,i_p$) yield
\begin{align}
\sum_{i_1 = i_2,\, i_3,\dots,i_p}
\xi_{i_1}^\mu \xi_{i_2}^\mu \cdots \xi_{i_p}^\mu \,
\sigma_{i_1} \sigma_{i_2} \cdots \sigma_{i_p}
= \sum_{i_1, i_3,\dots,i_p}
(\xi_{i_1}^\mu)^2 \xi_{i_3}^\mu \cdots \xi_{i_p}^\mu \,
(\sigma_{i_1})^2 \sigma_{i_3} \cdots \sigma_{i_p}.
\end{align}
Since $(\xi_i^\mu)^2 = 1$ and $(\sigma_i)^2 = 1$, such terms reduce to effective $(p-2)$-body interactions:
\begin{align}
\sim \sum_{i_1, i_3,\dots,i_p}
\xi_{i_3}^\mu \cdots \xi_{i_p}^\mu \,
\sigma_{i_3} \cdots \sigma_{i_p}.
\end{align}
More generally, contributions with $k$ repeated indices generate effective $(p-2k)$-body interaction terms.  
As a result, the expansion of $\left( \sum_i \xi_i^\mu \sigma_i \right)^p$ produces not only genuine $p$-body interactions but also a hierarchy of lower-order effective interactions.

These diagonal contributions are absent in the Abbott--Arian-type formulation, where the Hamiltonian is explicitly constructed from products of distinct indices.  
Our primary objective is to examine whether this difference gives rise to qualitative distinctions between the two models.

\section{Equilibrium storage capacity (replica result, for context)}
The equilibrium storage capacity of the $p$-body Hopfield model has been studied via replica theory.
For completeness and to set a baseline for the dynamical analysis, we summarize the zero-temperature fixed-point equation for the overlap with the retrieved pattern 
using probabilists’ Hermite polynomial.

\subsection{Diagonal-term identity and probabilists' Hermite polynomials}
When separating the Hamiltonian into the signal ($\mu=0$) and crosstalk noise ($\mu\ge 1$), diagonal contributions appear in the $p$-body sums.
Introduce
\begin{align}
\begin{cases}
    m = \frac{1}{N}\sum_{i} \xi_i^0 \sigma_i
    &\text{(overlap with pattern $\mu=0$)},\\[4pt]
    u_\mu = \frac{1}{\sqrt{N}}\sum_{i} \xi_i^\mu \sigma_i
    &\text{(crosstalk variable for pattern $\mu$)}.
\end{cases}
\label{eq:def_m_u}
\end{align}
Then, for $\mu\ge 1$,
\begin{align}
    &\sum_{j_1< j_2 < j_3 \cdots< j_p}
    \frac{\xi_{j_1}^\mu \sigma_{j_1}}{\sqrt{N}}
    \frac{\xi_{j_2}^\mu \sigma_{j_2}}{\sqrt{N}}
    \cdots
    \frac{\xi_{j_p}^\mu \sigma_{j_p}}{\sqrt{N}}
    \nonumber\\
    &= \sum_{d=0}^{\left\lfloor\frac{p}{2}\right\rfloor}
    (2d-1)!!\, (-1)^d \binom{p}{2d}\, u_{\mu}^{p-2d}
    + \mathcal{O}\!\left(\frac{1}{\sqrt{N}}\right)
    \nonumber\\
    &= \He_p(u_\mu) + \mathcal{O}\!\left(\frac{1}{\sqrt{N}}\right),.
    \label{eq:He_identity}
\end{align}
Here $\He_n(x)$ denotes the probabilists' Hermite polynomial,
\begin{align}
\He_n(x) := (-1)^n e^{x^2/2}\frac{\mathrm{d}^n}{\mathrm{d}x^n}\left(e^{-x^2/2}\right),
\end{align}
satisfying $\He_{n+1}(x)=x\,\He_n(x)-n\,\He_{n-1}(x)$.

\subsection{Saddle-point equation}
Using \eqref{eq:He_identity}, the partition function can be approximated as
\begin{align}
    Z \approx \sum_{\boldsymbol{\sigma}}
        \exp\left[
            \beta \left(
                \frac{N m^{p}}{p!}
                + \frac{1}{p!N^{\frac{p-2}{2}}} \sum_{\mu \ge 1}^{M-1}
                    \He_p(u_{\mu})
            \right)
        \right].
\end{align}
Carrying out the replica calculation under the replica symmetric ansatz leads to the self-consistent equation \cite{Gardner1987,Abbott1987}
\begin{align}
    m
    =
    \mathrm{erf}\!\left(
        \frac{m^{p-1}}{\sqrt{2 \alpha (p-1)! }}
    \right).
    \label{eq:capacity_fixedpoint}
\end{align}
in the zero-temperature limit.
We will compare dynamical basins with the critical loading $\alpha_c(p)$ implied by \eqref{eq:capacity_fixedpoint}.

\section{Retrieval dynamics from dynamical mean-field theory (DMFT)}

We analyze the synchronous dynamics in Eqs.~\eqref{eq:update}--\eqref{eq:field} in the large-$N$ limit using a path-integral formulation of dynamical mean-field theory (DMFT), focusing on the zero-temperature limit $\beta \to \infty$~\cite{DeDominicis1978,Coolen2000,Kabashima2026}.

\subsection{Generating functional}

To this end, we introduce source fields $\ell_i(t)$ and define the generating functional as
\begin{align}
\begin{split}
    \mathbb{E}_{\bm{J}}[Z(\tilde{\bm{\ell}} |\boldsymbol{J})]
    &=
    \mathop{\rm tr}_{\bm{h}(\cdot), \bm{\sigma}(\cdot)}
    P(\bm{\sigma}(0))
    \prod_{i=1}^N \prod_{t=0}^{T-1}
    \Big\{
        \delta(\sigma_i(t+1) - \sgn(h_i(t)))\, e^{i \ell_i(t) \sigma_i(t)}
    \Big\}
    \\
    &\qquad\times
    \mathbb{E}_{\bm{J}}\left[
    \prod_{i=1}^N \prod_{t=0}^{T-1}
    \delta\!\left(
        h_i(t) - \sum_{j_2 < j_3 \cdots< j_p} J_{i,j_2,\cdots,j_p}\,
        \sigma_{j_2}(t)\cdots\sigma_{j_p}(t)
    \right)
    \right]  \\
    &=\frac{1}{(2\pi)^{NT}} \mathop{\rm tr}_{\bm{h}(\cdot), \bm{\sigma}(\cdot), \bm{g}(\cdot)}
    \prod_{i=1}^N P(\sigma_i(0))\prod_{t=0}^{T-1}
    \Big\{
        \delta(\sigma_i(t+1) - \sgn(h_i(t)))\, e^{i \ell_i(t) \sigma_i(t)}
    \Big\}
    \\
    &\qquad\times
    \mathbb{E}_{\bm{J}}\left[
    \prod_{i=1}^N \prod_{t=0}^{T-1}
    \exp\!\left(
    ig_i(t)\left (
        h_i(t) - \sum_{j_2 < j_3 \cdots< j_p} J_{i,j_2,\cdots,j_p}\,
        \sigma_{j_2}(t)\cdots\sigma_{j_p}(t)
        \right )
    \right)
    \right], 
\end{split}
\label{eq:GF}
\end{align}
Here, $P(\bm{\sigma}(0))=\prod_{i=1}^N P(\sigma_i(0))$ is the distribution of 
$\bm{\sigma}(0) \in \{+1,-1\}^N$ and $\tilde{\bm{\ell}}=(\ell(0), \ell(1), \ldots, \ell(T))$. The notation $\mathbb{E}_X[\cdots]$ denotes an average over $X$, and $\mathrm{tr}_X\{\cdots\}$ represents summation or integration over all configurations of $X$.

The generating functional satisfies
\begin{align}
\lim_{\tilde{\bm{\ell}}\to \bm{0}} \mathbb{E}_{\bm{J}}[Z(\tilde{\bm{\ell}}|\boldsymbol{J})] &= 1, \cr
\lim_{\tilde{\bm{\ell}}\to \bm{0}} \frac{\partial}{\partial (i \ell_i(t))} \mathbb{E}_{\bm{J}}[Z(\tilde{\bm{\ell}}|\boldsymbol{J})] &= \mathbb{E}_{\bm{\sigma}(\cdot), \boldsymbol{J}}[\sigma_i(t)], \cr 
\lim_{\tilde{\bm{\ell}}\to \bm{0}} \frac{\partial^2}{\partial (i \ell_i(t)) \partial (i \ell_j(s))} \mathbb{E}_{\bm{J}}[Z(\tilde{\bm{\ell}}|\boldsymbol{J})] &= \mathbb{E}_{\bm{\sigma}(\cdot), \boldsymbol{J}}[\sigma_i(t)\sigma_j(s)]. \cr
& \cdots \nonumber
\end{align}
Hence, statistical properties of the retrieval dynamics can be directly obtained from the generating functional in Eq.~\eqref{eq:GF}.

\subsection{Order parameters and saddle-point relations}
The main difficulty in evaluating Eq.~\eqref{eq:GF} lies in computing
\begin{align}
\Xi &= \mathbb{E}_{\bm{J}}\left[
    \prod_{i=1}^N \prod_{t=0}^{T-1}
    \exp\!\left(
    ig_i(t)\left (
        h_i(t) - \sum_{j_2 < j_3 \cdots< j_p} J_{i,j_2,\cdots,j_p}\,
        \sigma_{j_2}(t)\cdots\sigma_{j_p}(t)
        \right )
    \right)
    \right]\cr
    &= \mathbb{E}_{\{ \bm{\xi}^\mu \}_{\mu=0}^{M-1}}
    \left[
    \exp\!\left(
    \sum_{i=1}^N\sum_{t=0}^{T-1} \!
    ig_i(t) \! \left (\! 
        h_i(t) \!-\! \frac{1}{N^{p-1} }
        \sum_{\mu=0}^{M-1} 
        \xi_i^\mu \!\!\!\!\!
        \sum_{j_2 < j_3 <\cdots< j_p} 
        \!\!\!\!\!
        (\xi_{j_2}^\mu \sigma_{j_2}(t))\cdots (\xi_{j_p}^\mu \sigma_{j_p}(t))
        \right )
    \right)
    \right]. 
    \end{align} 
Assuming without loss of generality that the retrieved pattern is $\bm{\xi}^0$, 
we introduce
\begin{align}
m(t) &= \frac{1}{N} \sum_{i=1}^N \xi_i^0 \sigma_i(t) = \mathcal{O}(1). 
\end{align}
On the other hand, we assume another scaling 
for noise patterns indexed by $\mu=1,2,\ldots, M-1$, as 
\begin{align}
u_\mu(t) &= \frac{1}{\sqrt{N}} \sum_{i=1}^N \xi_i^\mu \sigma_i(t) = \mathcal{O}(1), \\
v_\mu(t) &= \frac{1}{\sqrt{N}} \sum_{i=1}^N \xi_i^\mu (i g_i(t)) = \mathcal{O}(1).
\end{align}
Using these definitions, the contributions from the retrieved pattern and noise patterns can be expressed as
 \begin{align}
 \text{(retrieved)}  &\quad  \frac{1}{N^{p-1}} \sum_{i=1}\sum_{t=0}^{T-1}(i\xi_i^0g_i(t))\sum_{j_2<j_3 <\cdots < j_p}
 (\xi_{j_2}^0 \sigma_{j_2}(t))\cdots (\xi_{j_p}^0 \sigma_{j_p}(t)) \cr
 &\quad \quad = \sum_{i=1}^N \sum_{t=0}^T ig_i (t) \frac{m(t)^{p-1}}{(p-1)!} + {\mathcal O}(T N^{-1}),  
 \label{eq:retrieval_contribution} \\
\text{(noise)}  & \quad \frac{1}{N^{p-1}} \sum_{i=1}\sum_{t=0}^{T-1} (i\xi_i^\mu g_i(t)) \sum_{j_2<j_3 <\cdots < j_p}
 (\xi_{j_2}^\mu \sigma_{j_2}(t))\cdots (\xi_{j_p}^\mu \sigma_{j_p}(t)) \cr
 &\quad \quad = \frac{1}{N^{\frac{p-1}{2}}}
\sum_{t=0}^T \He_{p-1}(u_\mu) v_\mu + {\mathcal O}(TN^{-\frac{p}{2}}).
\label{eq:noise_contribution}
 \end{align}

The following two observations are crucial for evaluating Eq.~\eqref{eq:GF}:
\begin{itemize}
\item The noise patterns $\bm{\xi}^2, \ldots, \bm{\xi}^{M-1}$ are independent Rademacher random variables. Therefore, their contribution can be evaluated for a single pattern and then raised to the power $M-1$.
\item As noted above, we drop the pattern index $\mu$ for the noise patterns in what follows. For fixed sets $\{\sigma_i(t)\}$ and $\{g_i(t)\}$, the central limit theorem indicates that the variables $u(t)$ and $v(t)$ follow a zero-mean multivariate Gaussian distribution with covariance
\begin{align}
\left \{
\begin{aligned}
\mathbb{E}_{\bm{\xi}}[u(t)u(s)]&= \frac{1}{N} \sum_{i=1}^N \sigma_i(t)\sigma_i(s) =: Q(t,s), \\
\mathbb{E}_{\bm{\xi}}[v(t)v(s)] &= \frac{1}{N} \sum_{i=1}^N (i g_i(t))(i g_i(s)) =: R(t,s), \\
\mathbb{E}_{\bm{\xi}}[u(t)v(s)] &= \frac{1}{N} \sum_{i=1}^N \sigma_i(t) (i g_i(s)) =: S(t,s).
\end{aligned}
\right .
\label{eq:order_params}
\end{align}
\end{itemize}

We employ these for evaluating $\Xi$ and substituting the following identities, 
\begin{align}
1 &= N \int \delta\left (\sum_{i=1}^N \sigma_i(t) \sigma_i(s) -N Q(t,s)  \right ){\mathrm d}Q(t,s) \cr
&=\frac{N}{2\pi} \int \exp\left [
\hat{Q}(t,s) \left (\sum_{i=1}^N \sigma_i(t) \sigma_i(s) -N Q(t,s)  \right ) \right ]{\mathrm d}\hat{Q}(t,s)
{\mathrm d}Q(t,s)
\label{eq:delta_constraint}
\end{align}
and similarly for $R(t,s)$ and $S(t,s)$ for $\forall{t},\forall{s} \in \{0,1,\ldots,T\}$, into the right hand side
of \eqref{eq:GF}. The normalization property $\lim_{\tilde{\bm{\ell}}\to \bm{0}} \mathbb{E}_{\bm{J}}[Z(\tilde{\bm{\ell}}|\boldsymbol{J})]=1$ for the addition of any external fields to Eq.~\eqref{eq:field} guarantees $R(t,s)=0$ for any pairs 
of $t$ and $s$. 
Evaluating integrals over $Q(t,s)$ and $S(t,s)$ using the saddle point method offers
\begin{align}
\left\{
\begin{aligned}
&\hat{Q}(t,s) = 0, \\
&\hat{R}(t,s) = \frac{\alpha}{(p-1)!} Q(t,s)^{p-1}, \\
&\hat{S}(t,s) = \frac{\alpha}{(p-2)!} Q(t,s)^{p-2} S(s,t). 
\end{aligned}
\right . 
\label{eq:hat_variables}
\end{align}

Substituting these results back into Eq.~\eqref{eq:GF} leads to the final expression of the generating functional (see Appendix~\ref{app:DMFT} for details):
\begin{align}
&\mathbb{E}_{\bm{J}} [Z(\tilde{\bm{\ell}} | \boldsymbol{J})] \simeq 
\prod_{i=1}^N 
\mathop{\rm tr}_{\bm{h}_i, \bm{\sigma}_i, \bm{\phi}_i} 
{\mathcal N}(\bm{\phi}_i \mid \bm{0}, \hat{R} )
P(\sigma_i(0)) \cr
&\times \prod_{t=0}^{T-1}
\left\{\!
\delta\!\left (\sigma_i(t+1)\!-\!\sgn(h_i(t))\! \right ) 
\delta\! \left(\!h_i(t)\!-\!\frac{\xi_i^0 m(t)^{p-1}}{(p-1)!}\!-\!\phi_i(t)\! + \!\sum_{s<t}\hat{S}(s, t)\sigma_i(s)\right)\! e^{i \ell_i(t)\sigma_i(t)}
\! \right\}, 
\label{eq:GF_final}
\end{align}
where ${\mathcal N}(\bm{z} \mid \bm{\mu}, \Sigma)$ 
denotes the multivariate Gaussian distribution with mean vector $\bm{\mu}$
and covariance matrix $\Sigma$, 
$\bm{h}_i = (h_i(0), \ldots, h_i(T-1))$, and similarly for $\bm{\sigma}_i$ and $\bm{\phi}_i$.

\subsection{Effective single-site process}
\label{subsec:single_site}
Equation~\eqref{eq:GF_final} implies that the macroscopic dynamics of Eqs.~\eqref{eq:update} and \eqref{eq:field} can be described by an effective single-site stochastic process:
\begin{align}
\left \{
\begin{aligned}
&h(t) = \frac{1}{(p-1)!} \xi^{0}m(t)^{p-1}  + \phi(t) - \sum_{s=0}^{t-1} \hat{S}(s,t)\, \sigma(s),\\[6pt]
&\sigma(t+1) = \sgn(h(t)),\\[6pt]
\end{aligned}
\right.
\label{eq:effective_process}
\end{align}
where $m(t)=\mathbb{E}_{\xi^0}[\xi^{0} \sigma(t)]$ and $\phi(t)$ is a zero-mean Gaussian process with covariance $\hat{R}(t,s)$.
In practice, Eq.~\eqref{eq:effective_process} can be solved numerically using a large number of samples ($N_{\rm s} \gg 1$), as described below.
\paragraph{Input}
$N_{\rm s}$, $\bm{\xi}^0 = (\xi_i^0) \in \{+1,-1\}^{N_{\rm s}}$, $\alpha \ge 0$, and $m(0) \in [-1,1]$.
\paragraph{Initialization}
Initialize $\bm{\sigma}(0) = (\sigma_i(0)) \in \{+1,-1\}^{N_{\rm s}}$ such that
\begin{align}
\frac{1}{N_{\rm s}} \bm{\xi}^0 \cdot \bm{\sigma}(0) = m(0).
\end{align}
Set $Q(0,0) = 1$, $\hat{R}(0,0) = \frac{\alpha}{(p-1)!}$, and $\hat{S}(0,0) = 0$.
\paragraph{Iteration}
For $t = 0,1,\ldots,T-1$, repeat the following steps:
\begin{itemize}
\item For each $i \in {1,\ldots,N_{\rm s}}$, draw a Gaussian random variable $\phi_i(t)$ with zero mean such that
\begin{align}
\mathbb{E}_{\bm{\phi}_i} [\phi_i(t)\phi_i(s)] = \hat{R}(t,s), \quad (0 \le s \le t).
\end{align}
This can be carried out by sampling 
\begin{align}
\phi_i(t) \sim {\mathcal N}\left (\phi_i(t) \mid -\frac{\sum_{s=0}^{t-1} \hat{R}^{-1}(t,s)\phi_i(s)}{\hat{R}^{-1}(t,t)}, 
\frac{1}{\hat{R}^{-1}(t,t)}\right ), 
\end{align}
where $\hat{R}^{-1}$ denotes the inverse of the matrix $\hat{R}=(R(a,b))_{0\le a,b \le t}$. 
\item Update the local field and neuron state:
\begin{align}
h_i(t) &= \frac{\xi_i^0m(t)^{p-1}}{(p-1)!} + \phi_i(t) - \sum_{s=0}^{t-1} \hat{S}(s,t)\sigma_i(s), \\
\sigma_i(t+1) &= \sgn(h_i(t)).
\end{align}
\item Compute the macroscopic observables:
\begin{align}
m(t+1) &= \frac{1}{N_{\rm s}} \sum_{i=1}^{N_{\rm s}} \xi_i^0 \sigma_i(t), \\
Q(t+1,s) &= \frac{1}{N_{\rm s}} \sum_{i=1}^{N_{\rm s}} \sigma_i(t)\sigma_i(s), \quad (0 \le s \le t+1),
\end{align}
and
\begin{align}
S(t+1,s) =
\begin{cases}
\displaystyle
-\sum_{u=0}^{t}
\left(
\frac{1}{N_{\rm s}} \sum_{i=1}^{N_{\rm s}} \sigma_i(t)\phi_i(u)
\right)
\hat{R}^{-1}(u,s),
& (0 \le s \le t), \\
0, & (s = t+1).
\end{cases}
\end{align}
For $t=0$, 
\begin{align}
S(1,0)
=- \sqrt{\frac{2(p-1)!}{\pi\alpha}}\,
   \exp\!\left( -\frac{m(0)^{2(p-1)}}{2(p-1)! \alpha} \right).
\label{eq:S10}
\end{align}
is analytically evaluated exceptionally. 
\item Using these quantities, update $\hat{R}(t+1,s)$ and $\hat{S}(s,t+1)$ for $s=0,1,\ldots,t+1$ according to Eq.~\eqref{eq:hat_variables}.
\end{itemize}

As a reference, we present the macroscopic dynamics of the Krotov--Hopfield model~\cite{Mimura2025}:
\begin{align}
\left\{
\begin{aligned}
&h(t) = p \xi^{0} m(t)^{p-1} + \phi(t) + \sum_{s=0}^{t-1} \Gamma(t,s) \sigma(s), \\[6pt]
&\sigma(t+1) = \sgn(h(t)),
\end{aligned}
\right.
\label{eq:Krotov_effective_process}
\end{align}
where $m(t) = \mathbb{E}_{\xi^0} [\xi^{0} \sigma(t)]$ and 
$\phi(t)$ is a zero-mean Gaussian process with covariance
\begin{align}
\mathbb{E}_{\bm{\phi}}[\phi(t)\phi(s)]
=  \alpha p^2 \sum_{k=0}^{p-1} A(p-1,k)(Q(t,s))^k.
\end{align}
The coefficients $A(\ell,k)$ are defined as
\begin{align}
A(\ell, k) =
\begin{cases}
 \binom{\ell}{k}^2k! \{(\ell - k-1)!!\}^2 , & \text{if }\ell -k  \text{ is even}, \\
0, & \text{otherwise},
\end{cases}
\end{align}
and the self-interaction kernel is given by
\begin{align}
\Gamma(t,s)
= \alpha p^2 (p-1)^2 \sum_{k=0}^{p-2} A(p-2,k) (Q(t,s))^k S(t,s).
\label{eq:Gamma}
\end{align}

The dynamical structure is analogous to Eq.~\eqref{eq:effective_process}. 
However, the inclusion of diagonal terms in the Hamiltonian leads 
to more intricate forms for both the noise covariance and the self-interaction kernel.

\section{Numerical results}

\subsection{Time evolution of the overlap}
We compare the DMFT prediction with direct simulations for a representative case ($p=3$, $\alpha=0.05 < \alpha_c(p=3)$).
The theory corresponds to the large-$N$ limit; simulations at modest $N$ can exhibit noticeable finite-size effects.

\begin{figure}[tbp]
  \centering
  \begin{minipage}{0.45\linewidth}
    \centering
    \includegraphics[width=\linewidth]{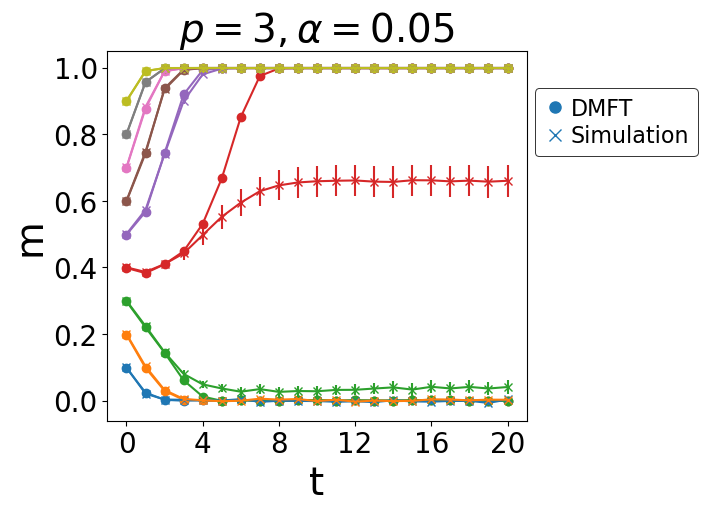}\\[-1.2em]
    (a) $\alpha=0.05$
  \end{minipage}\hfill
  \begin{minipage}{0.45\linewidth}
    \centering
    \includegraphics[width=\linewidth]{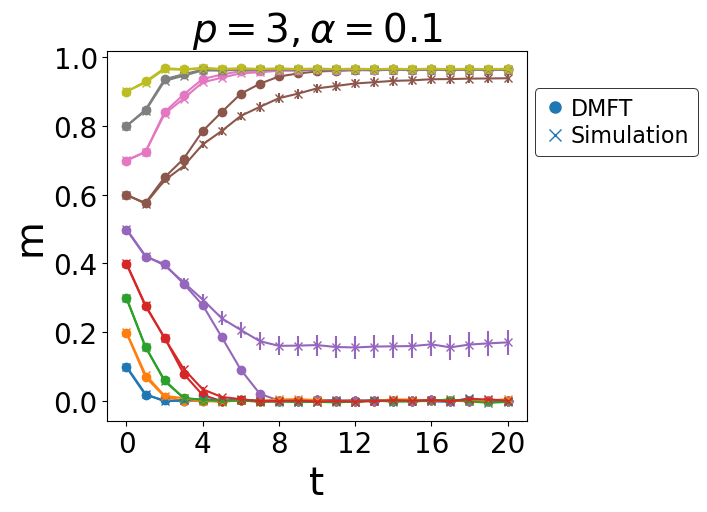}\\[-1.2em]
    (b) $\alpha=0.1$
  \end{minipage}

  \begin{minipage}{0.45\linewidth}
    \centering
    \includegraphics[width=\linewidth]{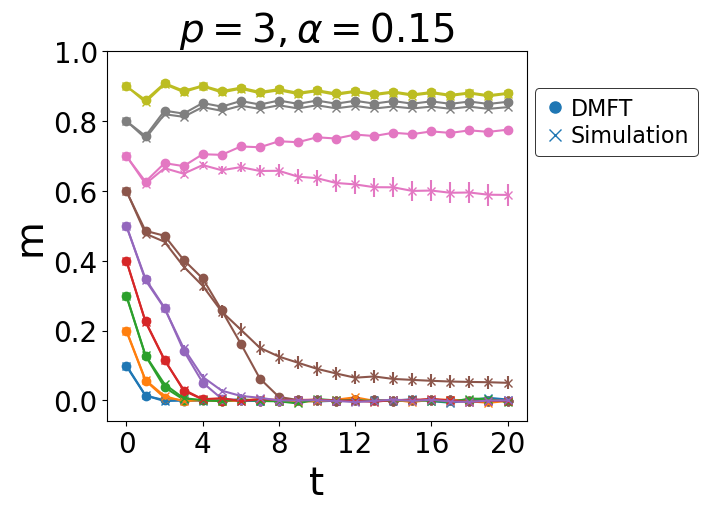}\\[-1.2em]
    (c) $\alpha=0.15$
  \end{minipage}\hfill
  \begin{minipage}{0.45\linewidth}
    \centering
    \includegraphics[width=\linewidth]{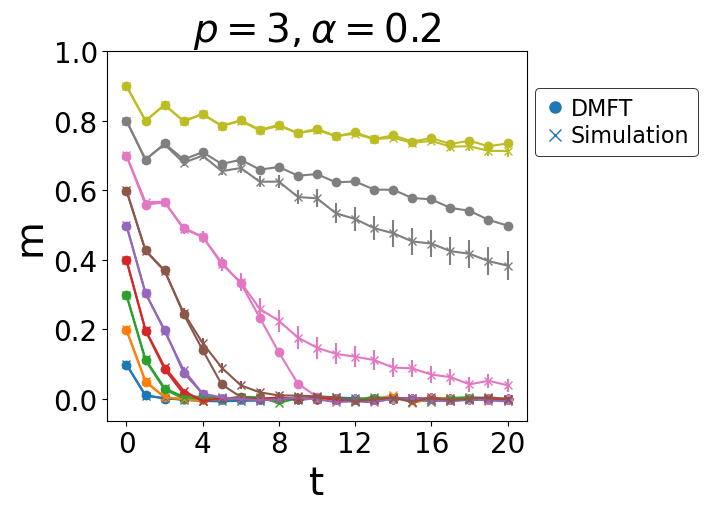}\\[-1.2em]
    (d) $\alpha=0.2$
  \end{minipage}
\caption{
Time evolution of the overlap $m(t)$ for $p=3$ and loading level $\alpha=0.05,0.1,0.15,0.2$.
Curves correspond to different initial overlaps $m(0)$.
Circles: DMFT prediction, obtained by solving the effective single-site equations with $N_{\rm s} = 
10^6$ Monte Carlo samples.
Crosses: direct simulations at $N=1024$ averaged over 100 runs; error bars show standard deviations.
Good agreement is observed over a wide time window
except for critical cases, which are presumably due to finite-size effects. 
}
  \label{fig:2x2}
\end{figure}
\FloatBarrier

\subsection{Basins of attraction}
We estimate basins of attraction by scanning the loading rate $\alpha$ and initial overlap $m(0)$, and recording the overlap after a finite number of iterations.
The yellow region corresponds to successful retrieval ($m(T)\approx 1$), while the purple region corresponds to convergence to the failure solution ($m(T)\approx 0$).
The intermediate-colored region, particularly in the upper-right part of each panel, should be interpreted as a finite-time non-converged regime: although $\alpha > \alpha_c$, the dynamics has not yet reached the failure fixed point $m=0$ within the observation time $T$.
The red dashed line indicates the critical loading $\alpha_c(p)$ from the equilibrium analysis \eqref{eq:capacity_fixedpoint}.

\begin{figure}[tbp]
  \centering
  \begin{minipage}{0.48\linewidth}
    \centering
    \includegraphics[width=\linewidth]{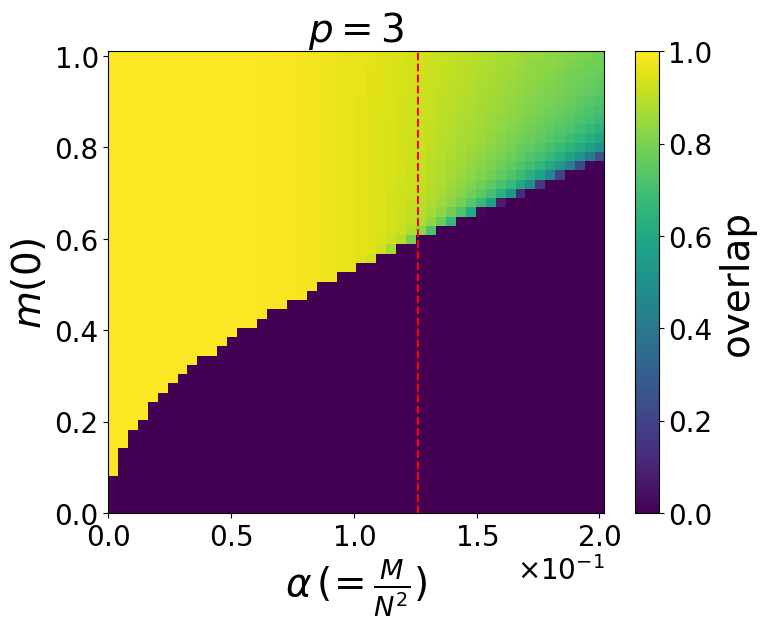}\\[-1.2em]
    (a) $p=3$
  \end{minipage}\hfill
  \begin{minipage}{0.48\linewidth}
    \centering
    \includegraphics[width=\linewidth]{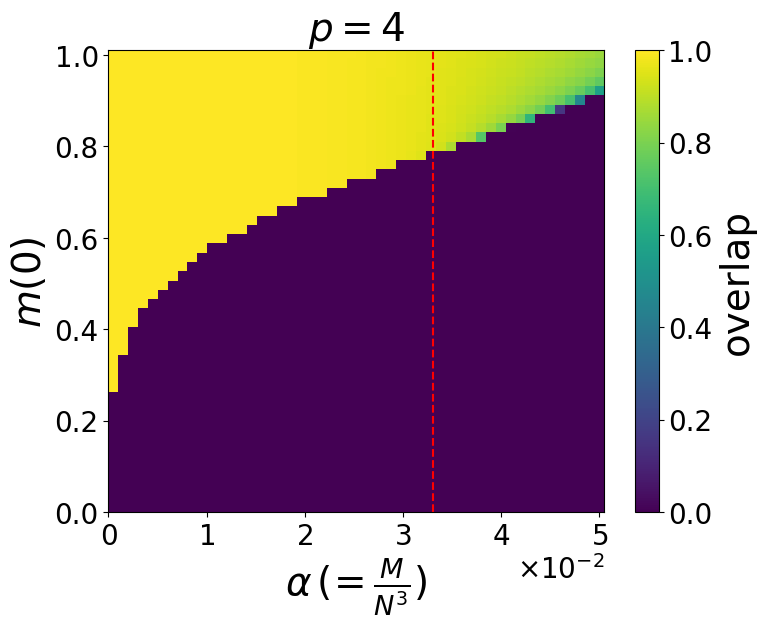}\\[-1.2em]
    (b) $p=4$
  \end{minipage}

  \vspace{0.5em}

  \begin{minipage}{0.48\linewidth}
    \centering
    \includegraphics[width=\linewidth]{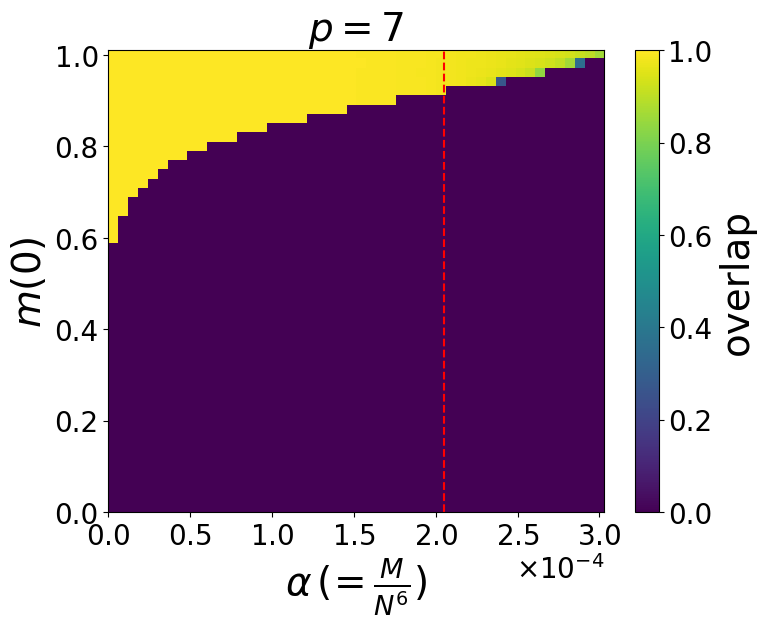}\\[-1.2em]
    (c) $p=7$
  \end{minipage}\hfill
  \begin{minipage}{0.48\linewidth}
    \centering
    \includegraphics[width=\linewidth]{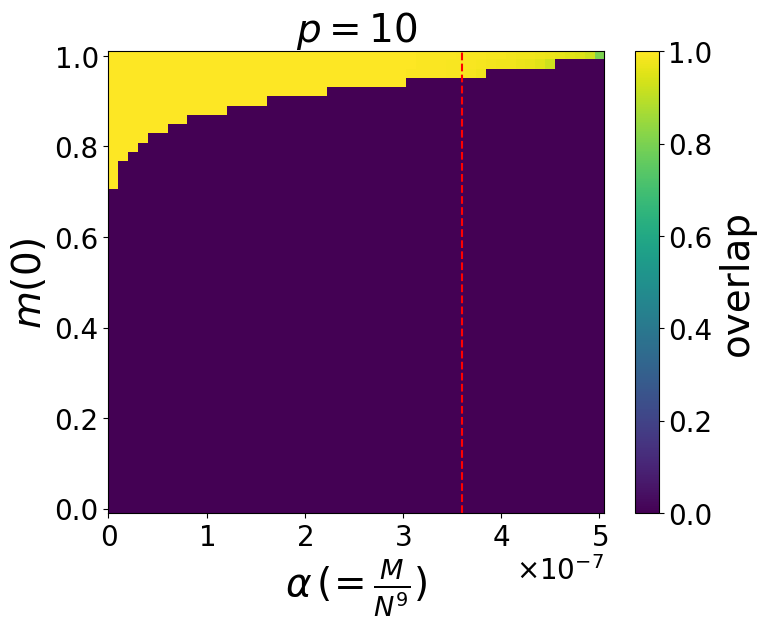}\\[-1.2em]
    (d) $p=10$
  \end{minipage}

\caption{
Finite-time retrieval performance as a function of the initial overlap $m(0)$ and loading level $\alpha$ for representative interaction orders $p=3,4,7,10$.
Colors represent the final overlap in runs with $T=20$, obtained from DMFT.
The gradual change of color indicates slow relaxation near the dynamical retrieval boundary.
The vertical dashed lines stand for the storage capacity evaluated by the RS static analysis. 
}
  \label{fig:2x2}
\end{figure}
\FloatBarrier

\begin{figure}[t]
  \centering

  \begin{subfigure}{0.95\textwidth}
    \centering
    \includegraphics[width=0.38\linewidth]{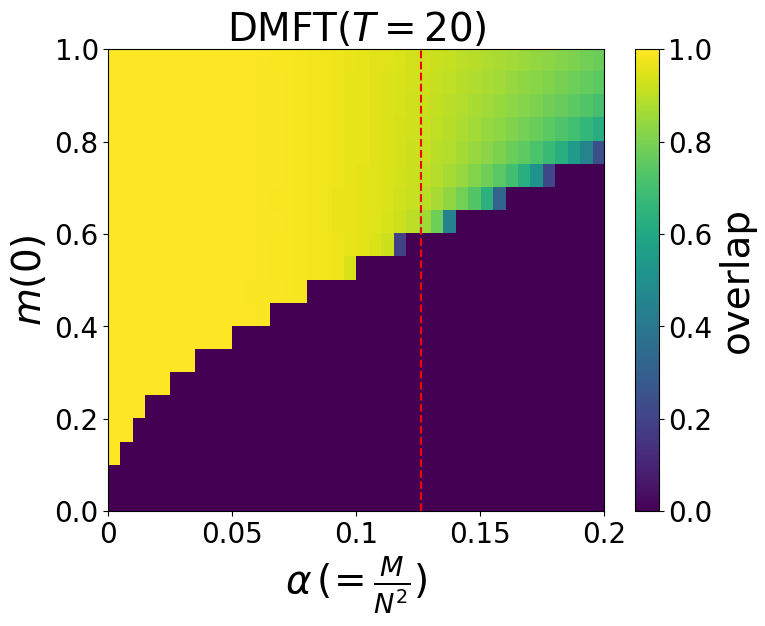}
    \includegraphics[width=0.38\linewidth]{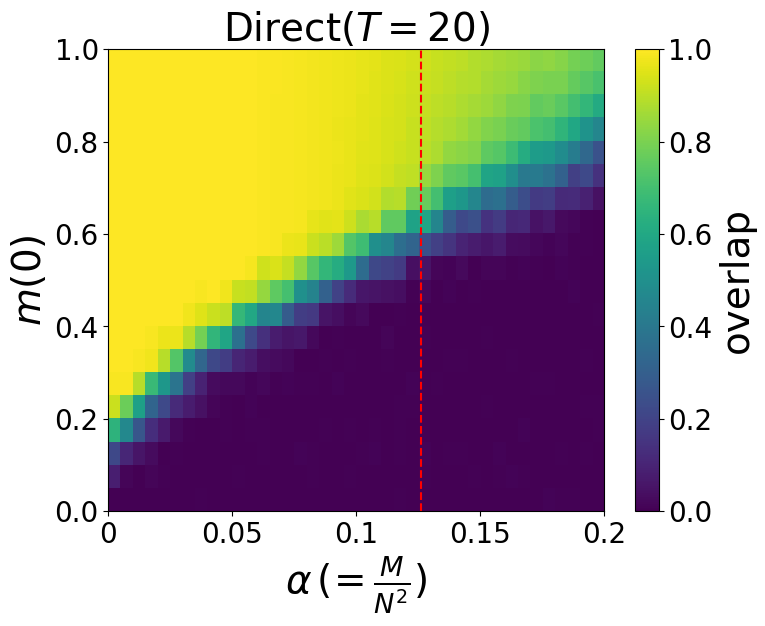}
    \vspace{-7mm}
    \caption{$T=20$}
    \label{fig:rowa}
  \end{subfigure}

  \begin{subfigure}{0.95\textwidth}
    \centering
    \includegraphics[width=0.36\linewidth]{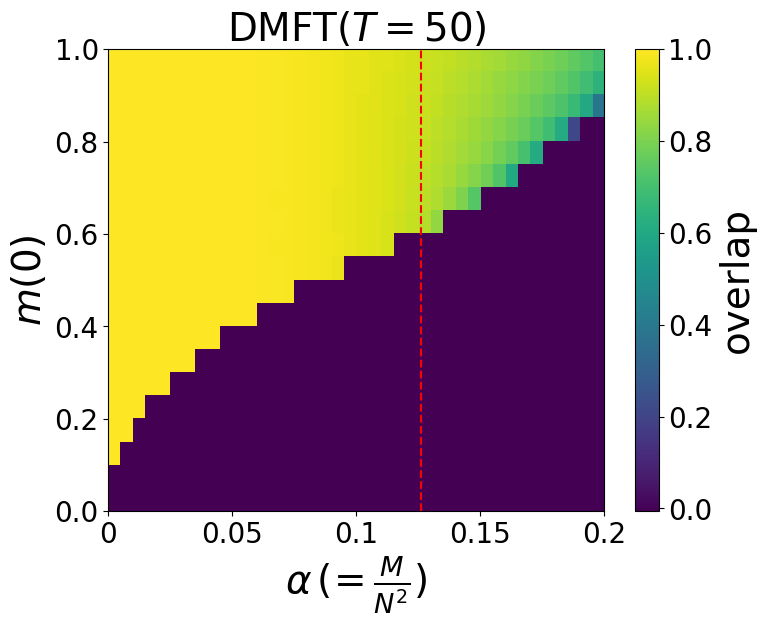}
    \includegraphics[width=0.36\linewidth]{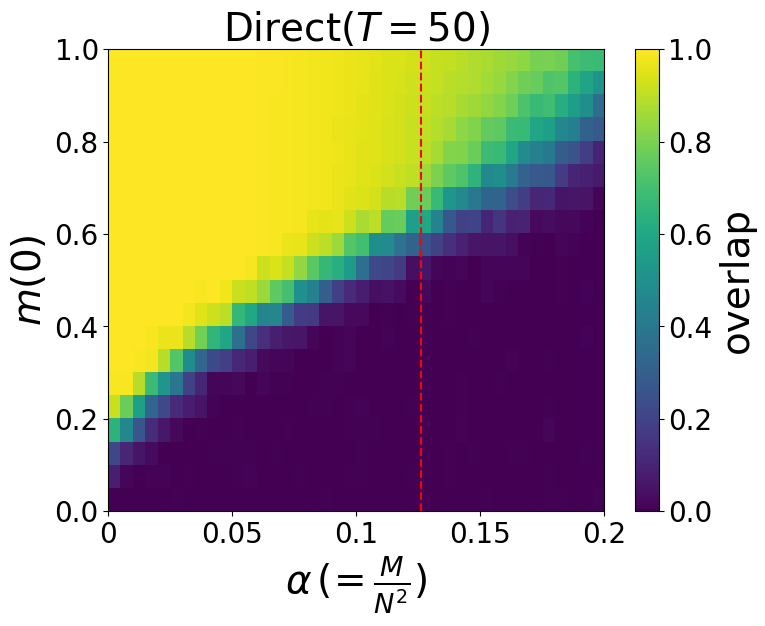}
    \vspace{-7mm}
    \caption{$T=50$}
    \label{fig:rowb}
  \end{subfigure}

  \begin{subfigure}{0.95\textwidth}
    \centering
    \includegraphics[width=0.36\linewidth]{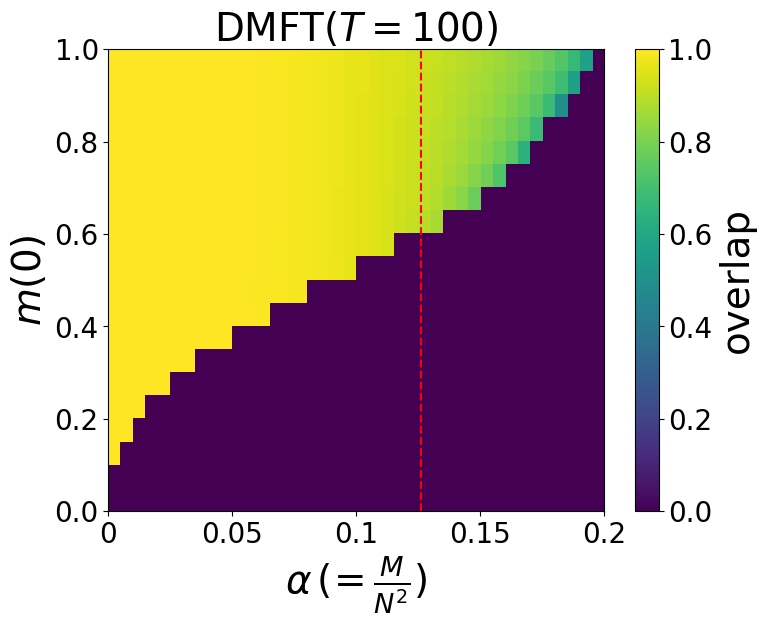}
    \includegraphics[width=0.36\linewidth]{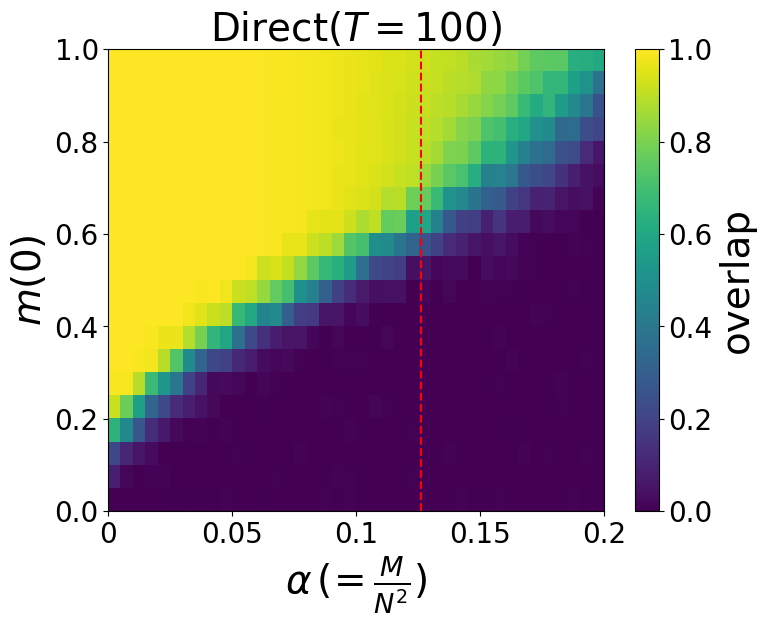}
    \vspace{-7mm}
    \caption{$T=100$}
    \label{fig:rowc}
  \end{subfigure}

  \begin{subfigure}{0.95\textwidth}
    \centering
    \includegraphics[width=0.36\linewidth]{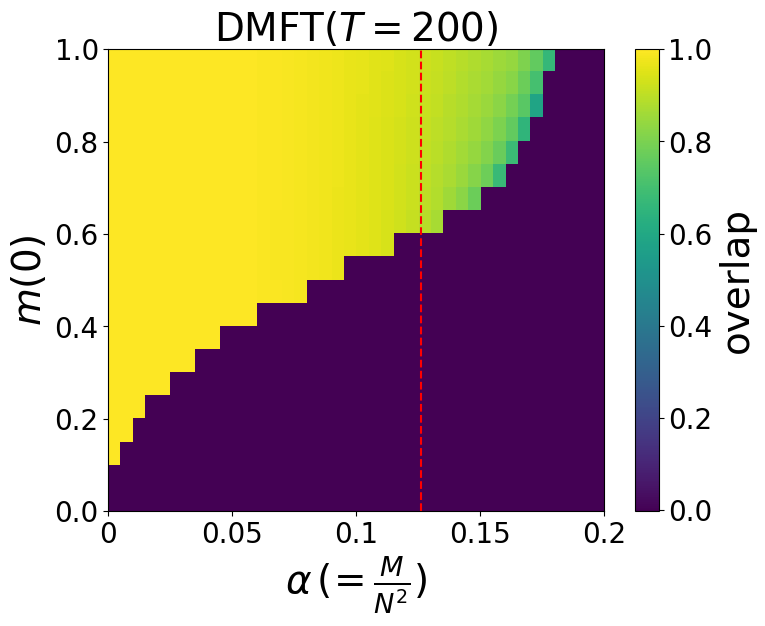}
    \includegraphics[width=0.36\linewidth]{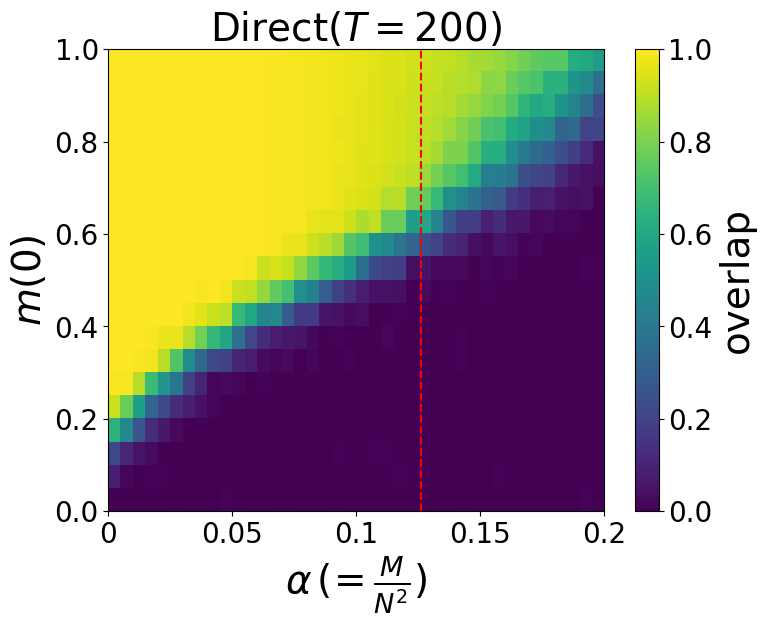}
    \vspace{-7mm}
    \caption{$T=200$}
    \label{fig:rowd}
  \end{subfigure}
\caption{
Overlap with the retrieved pattern 
after $T = 20,50,100,200$ iterations for
$p = 3$. Left: DMFT prediction, obtained by solving the effective single-site equations with $10^6$ Monte Carlo samples.
right: direct simulations at $N=1024$ averaged over 100 runs.
The vertical dashed lines stand for the storage capacity evaluated by the RS static theory. 
}
  \label{fig:direct_dmft_compare}
\end{figure}

We next compare the DMFT basin maps with direct simulations in the same spirit as the
side-by-side comparison in Ref.~\cite{Mimura2025}.
Figure~\ref{fig:direct_dmft_compare} shows the results for $T=20,50,100,200$, with DMFT in the left column and direct simulations in the right column.
The overall structure and its $T$ dependence agree well between the two approaches.
The direct simulations exhibit a smoother crossover near the boundary, which is attributable to
finite-size fluctuations and sample averaging, whereas DMFT describes the $N\to\infty$ limit.

\FloatBarrier

\begin{figure}[htbp]
  \centering
  \includegraphics[width=0.6\textwidth]{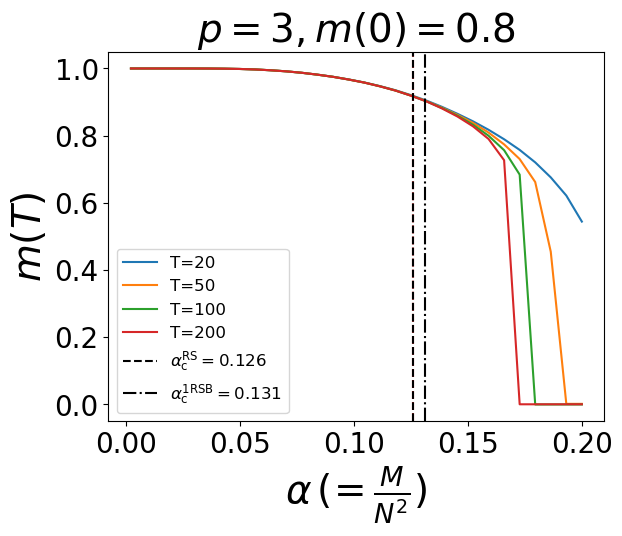}
\caption{
Final overlap $m(T)$ as a function of the loading rate $\alpha$ for $p=3$ with fixed initial overlap
$m(0)=0.8$.
Curves are shown for $T=20,50,100,200$.
As the observation time increases, the drop in $m(T)$ shifts toward smaller $\alpha$.
The vertical dashed and dash-dotted lines indicate the critical loadings from the RS and 1RSB
static analyses, respectively.
}
  \label{fig:mT_alpha}
\end{figure}

To further clarify the interpretation of the intermediate-colored region, Fig.~\ref{fig:mT_alpha}
shows the finite-time dependence of the final overlap for fixed initial overlap $m(0)=0.8$ in the
$p=3$ case.
The vertical lines indicate both static reference values, namely the replica-symmetric estimate
$\alpha_c^{\mathrm{RS}}$ and the 1-step replica-symmetry-breaking estimate $\alpha_c^{\mathrm{1RSB}}$.
Even for $\alpha>\alpha_c^{\mathrm{RS}}$, and in part even beyond $\alpha_c^{\mathrm{1RSB}}$, the overlap $m(T)$ can remain positive when the number of updates is
finite, indicating that the trajectory has not yet reached the failure fixed point $m=0$.
As $T$ increases, the drop in $m(T)$ shifts toward smaller $\alpha$, supporting the interpretation
that the intermediate-colored region is largely a finite-time non-converged regime rather than a
distinct stable retrieval phase.

\FloatBarrier

As in the case of the Krotov--Hopfield model, 
near the retrieval/failure boundary, the DMFT iteration exhibits very slow relaxation:
increasing the iteration count from 20 to 200 does not show a clear tendency toward
convergence in part of the parameter region.
Moreover, even beyond the storage limit predicted by the replica-symmetric (RS) theory, the overlap can remain substantially positive over long but finite times, reflecting slow relaxation rather than genuine stability of the retrieval state.
In fact, it has been shown that the RS assumption is violated in this system \cite{Abbott1987}.
However, even when the 1-step replica-symmetry-breaking (1RSB) estimate $\alpha_c^{\mathrm{1RSB}}(3)\simeq 0.131$ is also taken into account, the discrepancy from the DMFT result for $T=200$ remains large.

Taken together, these observations suggest that near the transition, the local energy landscape around the retrieved state is highly rugged and exhibits glassy features, including long-lived metastable trapping, similar to those observed in the Krotov--Hopfield model.
Clarifying this point will likely require analyses of further RSB steps,
including longer-time dynamics and/or treatments that incorporate RSB effects.
At the same time, extending to much longer times is computationally demanding in our
current implementation, since the DMFT iteration scales as $\mathcal{O}(T^3)$ with the
time horizon $T$.

\section{Conclusion}
We investigated the retrieval dynamics of the binary Abbott--Arian-type $p$-body Hopfield model using dynamical mean-field theory (DMFT) in the large-$N$ limit at zero temperature.  
The main objective of this study was to examine whether the slow retrieval dynamics observed in high-order associative memory models originates from diagonal (self-interaction) contributions, as present in the Krotov--Hopfield-type dense associative memory model.

To this end, we analyzed the Abbott--Arian-type model, in which such diagonal contributions are absent by construction.  
Using DMFT, we derived an effective single-site process together with closed macroscopic equations governing the overlap dynamics.  
In this derivation, probabilists' Hermite polynomials provide a convenient representation of the $p$-body interaction terms, enabling a systematic treatment of the dynamics.

The DMFT predictions are in good agreement with direct numerical experiments.  
At the same time, even with substantial computation time, the retrieval success/failure boundary obtained dynamically deviates markedly from the storage capacity predicted by static analysis.  
In particular, the apparent basin of attraction observed dynamically extends well beyond that predicted by equilibrium statistical mechanics.  

Importantly, these features persist even in the absence of diagonal self-interaction terms.  
This demonstrates that the slow, glass-like relaxation near the retrieval boundary cannot be attributed primarily to such diagonal contributions, but instead originates from intrinsic properties of high-order interactions.  
This suggests that the slow dynamics is associated with a rugged effective energy landscape induced by higher-order interactions.

An important direction for future work is to clarify how static analyses incorporating replica symmetry breaking can be reconciled with DMFT in describing this slow, glassy relaxation.
It would also be interesting to investigate whether the present mechanism extends to broader classes of high-order neural network models.

\section*{Acknowledgements}
The authors would like to thank Kazushi Mimura for fruitful discussions. 
This work was supported in part by MEXT/JSPS KAKENHI Grant Number 22H05117 (YK).

\appendix
\section{DMFT derivation}
\label{app:DMFT}
This appendix summarizes the path-integral derivation that leads from the microscopic dynamics \eqref{eq:update}--\eqref{eq:field} to the effective single-site process \eqref{eq:effective_process}. The treatment follows standard DMFT derivations (e.g.\ \cite{DeDominicis1978,Coolen2000}), with a key model-specific ingredient being the exact subtraction of diagonal contributions in the $p$-body crosstalk term via probabilists' Hermite polynomials.

\subsection{Fourier representation and signal/noise separation}
Starting from the generating functional \eqref{eq:GF}, we Fourier-transform the dynamical constraint
for each $(i,t)$:
\begin{align}
\delta\!\Bigl(
h_i(t)
\!-\! \sum_{j_2<\cdots<j_p} J_{i,j_2,\dots,j_p}
\,\prod_{\ell=2}^{p}\sigma_{j_\ell}(t)
\Bigr)
\!=\!
\int \frac{\mathrm{d}g_i(t)}{2\pi}
\exp\!\left[
i g_i(t)\!\Bigl(
h_i(t)
\!-\! \sum_{j_2<\cdots<j_p} J_{i,j_2,\dots,j_p}
\,\prod_{\ell=2}^{p}\sigma_{j_\ell}(t)
\Bigr)
\right].
\label{eq:app_fourier}
\end{align}
Substituting \eqref{eq:J}, the interaction contribution in the exponent can be written as
\begin{align}
-\frac{i}{N^{p-1}}\sum_{t}\sum_{\mu=0}^{M-1}\sum_{i} g_i(t)\,\xi_i^\mu
\sum_{j_2<\cdots<j_p}
\prod_{\ell=2}^{p}\xi_{j_\ell}^\mu\sigma_{j_\ell}(t).
\label{eq:app_signal_noise_split}
\end{align}
We then separate the condensed pattern $\mu=0$ (signal) from the non-condensed patterns $\mu\ge 1$
(crosstalk). For the signal part ($\mu=0$), the sum over $(j_2,\dots,j_p)$ with diagonal exclusions yields, to leading
order in $N$,
\begin{align}
\sum_{j_2<\cdots<j_p}
\xi_{j_2}^0\sigma_{j_2}(t)\cdots \xi_{j_p}^0\sigma_{j_p}(t)
=
\frac{N^{p-1}}{(p-1)!}\,m(t)^{p-1}
+ \mathcal{O}(N^{p-2}),
\label{eq:app_signal_leading}
\end{align}
so that the condensed contribution is $-(i/(p-1)!)\sum_{i,t} g_i(t)\,\xi_i^0\,m(t)^{p-1}$.

\subsection{Diagonal-excluded crosstalk noises and Hermite polynomials}
For the crosstalk noise parts ($\mu\ge 1$), introduce for each non-condensed pattern
\begin{align}
u_\mu(t) := \frac{\boldsymbol{\xi}^\mu\!\cdot\!\boldsymbol{\sigma}(t)}{\sqrt{N}},
\qquad
v_\mu(t) := \frac{\boldsymbol{\xi}^\mu\!\cdot\! i\boldsymbol{g}(t)}{\sqrt{N}}.
\label{eq:app_uv_def}
\end{align}
The diagonal-excluded $(p\!-\!1)$-fold sum admits the asymptotic identity (for $\mu\ge 1$)
\begin{align}
\sum_{j_2 < j_3 \cdots< j_p}
\frac{\xi_{j_2}^\mu\sigma_{j_2}(t)}{\sqrt{N}}\cdots
\frac{\xi_{j_p}^\mu\sigma_{j_p}(t)}{\sqrt{N}}
=
\He_{p-1}\!\big(u_\mu(t)\big) + \mathcal{O}\!\left(\frac{1}{\sqrt{N}}\right),
\label{eq:app_He_dyn}
\end{align}
which is the dynamical counterpart of \eqref{eq:He_identity} (note the order $p-1$ here because the local
field involves $p-1$ other spins). Therefore, the non-condensed contribution can be written as
\begin{align}
\exp\!\left\{
-\frac{1}{N^{\frac{p-2}{2}}(p-1)!}
\sum_{\mu\ge 1}\sum_{t} \He_{p-1}\!\big(u_\mu(t)\big)\,v_\mu(t)
\right\}.
\label{eq:app_noise_factor}
\end{align}

\subsection{Noise average, quadratic truncation, and macroscopic order parameters}

Let $\Xi$ denote the average over the non-condensed patterns $\{\boldsymbol{\xi}^\mu\}_{\mu \ge 1}$ of Eq.~\eqref{eq:app_noise_factor}. 
Using the independence across $\mu$ and $\alpha = M/N^{p-1}$, we obtain
\begin{align}
\Xi
&=
\mathbb{E}_{\bm{\xi}}\left[
\exp\left(
-\frac{1}{N^{\frac{p-2}{2}}(p-1)!}\sum_{t} \He_{p-1}\big(u(t)\big)\, v(t)
\right)
\right]^{\alpha N^{p-1}-1},
\label{eq:app_Xi_single}
\end{align}
where $u(t) = (\boldsymbol{\xi} \cdot \boldsymbol{\sigma}(t))/\sqrt{N}$ and 
$v(t) = (\boldsymbol{\xi} \cdot i\boldsymbol{g}(t))/\sqrt{N}$ for a single random pattern $\boldsymbol{\xi}$.

For $p \ge 3$, we have $\mathbb{E}_{\bm{\xi}}[He_{p-1}(u(t))v(t)] = 0$. 
Expanding the exponential up to second order (higher-order terms are subleading in $N$ under the present scaling), we obtain
\begin{align}
&\Xi \approx
\exp\left(
\frac{N\alpha}{2((p-1)!)^2}
\sum_{t,s}
\mathbb{E}_{\bm{\xi}}\big[
\He_{p-1}(u(t))\He_{p-1}(u(s))\, v(t)v(s)
\big]
\right).
\label{eq:app_Xi_quad}
\end{align}

To evaluate the remaining expectation, we use the fact that, by the central limit theorem, the process 
$\{(u(t), v(t))\}_{t=0}^{T}$ is asymptotically jointly Gaussian with covariances given by Eq.~\eqref{eq:order_params}. 
This enables us to apply Stein's lemma
\begin{align}
\mathbb{E}_{\bm{x}} \left[ x_i f(x_1,\ldots,x_n) \right]
=
\sum_{j=1}^n \mathbb{E}_{\bm{x}}[x_i x_j]\,
\mathbb{E}_{\bm{x}} \left[\frac{\partial f(x_1,\ldots,x_n)}{\partial x_j} \right],
\label{eq:Stein}
\end{align}
valid for any differentiable function $f$, together with standard identities for probabilists' Hermite polynomials \cite{Olver2010NIST},
\begin{align}
\He_n'(x) = n\,\He_{n-1}(x), \qquad
\mathbb{E}_{x,y}\big[\He_m(x)\He_n(y)\big]
= \delta_{mn}\, m!\, \mathbb{E}_{x,y}[xy]^m,
\label{eq:He_formulas}
\end{align}
where the second identity holds for jointly Gaussian variables with unit variances.

Applying Stein's lemma sequentially to $v(t)$ and $v(s)$, in conjunction with 
Eq.~\eqref{eq:He_formulas}, 
we obtain
\begin{align}
&\mathbb{E}\sqbra{\He_{p-1}(u(t))\He_{p-1}(u(s))\,v(t)v(s)}\cr
&=\mathbb{E}[v(t)u(t)]\mathbb{E}[\He_{p-1}^\prime(u(t)) \He_{p-1}(u(s)) v(s) ]
+ \mathbb{E}[v(t)u(s)]\mathbb{E}[\He_{p-1}(u(t)) \He_{p-1}^\prime(u(s)) v(s) ] \cr
& \hspace{2cm} +\mathbb{E}[v(t)v(s)] \mathbb{E}[\He_{p-1}(u(t))\He_{p-1}(u(s))]
\cr
&= \mathbb{E}[v(t)u(t)] \mathbb{E}[v(s)u(s)]\mathbb{E}[\He_{p-1}^\prime(u(t))\He_{p-1}^\prime(u(s))]\cr
& \hspace{2cm} + \mathbb{E}[v(t)u(s)]\mathbb{E}[v(s)u(t)]\mathbb{E}[\He_{p-1}^\prime (u(t)) \He_{p-1}^\prime(u(s)) ] \cr
& \hspace{2cm} + \mathbb{E}[v(t)v(s)] \mathbb{E}[\He_{p-1}(u(t))\He_{p-1}(u(s))] \cr
&=(p-1)^2(p-2)!S(t,t)S(s,s)Q(t,s)^{p-2} +(p-1)^2(p-2)! S(t,s)S(s,t) (p-2)! Q(t,s)^{p-2} \cr
&\hspace{2cm} + R(t,s) (p-1)! Q(t,s)^{p-1} \cr
&= (p-1)!\Big\{Q(t,s)^{p-1}R(t,s)
+(p-1)Q(t,s)^{p-2}\big(S(t,t)S(s,s)
+S(t,s)S(s,t)\big)
\Big\}, 
\label{eq:app_He_contraction}
\end{align}
where we abbreviated $\mathbb{E}_{\bm{u},\bm{v}}$ as $\mathbb{E}$.

Substituting this into Eq.~\eqref{eq:app_Xi_quad} yields
\begin{align}
&\Xi(Q,S,R) \cr
&\approx
\exp\left(
\frac{N\alpha}{2(p-1)!}
\sum_{t,s}\!
\Big[\! 
Q(t,s)^{p-1}R(t,s)
\!+\! (p\!-\!1)Q(t,s)^{p-2}
\big(
\!S(t,t)S(s,s) \!+\! S(t,s)S(s,t)
\! \big)
\! \Big]
\right).
\label{eq:app_Xi_final}
\end{align}

Combining this with the contributions from Eq.~\eqref{eq:delta_constraint} and those involving $R$ and $S$, we finally obtain
\begin{align}
\frac{1}{N} \ln \mathbb{E}_{\bm{J}} \left[Z(\tilde{\bm{\ell}}|\bm{J})\right]
&=
\mathrm{extr}_{\{Q,S,R,\hat{Q},\hat{S},\hat{R}\}}
\Bigg\{
\ln \Xi(Q,S,R)
+ \frac{1}{N}\sum_{i=1}^N \ln \Omega(\bm{\ell}_i; \hat{Q},\hat{S},\hat{R},\xi_i^0)
\nonumber \\
&\qquad
- \frac{1}{2} \sum_{t,s}
\big(
\hat{Q}(t,s)Q(t,s)
+ 2\hat{S}(t,s)S(t,s)
+ \hat{R}(t,s)R(t,s)
\big)
\Bigg\},
\label{eq:free_energy}
\end{align}
where $\mathop{\rm extr}_X\{\cdots\}$ denote extremization with respect to $X$ and 
\begin{align}
&\Omega(\bm{\ell}_i;\hat{Q}, \hat{S}, \hat{R}, \xi_i^0)\cr
&=\mathop{\rm tr}_{\bm{h}_i, \bm{\sigma}_i, \bm{g}_i}
\left \{
P(\sigma_i(0))\prod_{t=0}^{T-1} \left (
\delta(\sigma_i(t+1)-\sgn(h_i(t))) e^{i\ell_i(t)\sigma_i(t)} \right )\right . \cr
&\quad \left . \times \exp\left (\frac{1}{2}(i\bm{g}_i)^\top \hat{R} (i\bm{g}_i) 
+\bm{\sigma}_i \hat{S} (i\bm{g}_i) + \frac{1}{2} \bm{\sigma}_i \hat{Q} \bm{\sigma}_i
+i\sum_{t=0}^{T-1}g_i(t) \left (h_i(t) -\frac{\xi_i^0 m(t)^{p-1}}{(p-1)!} \right ) \right ) \right \}.  \cr
&=\mathop{\rm tr}_{\bm{h}_i, \bm{\sigma}_i, \bm{g}_i, \bm{\phi}_i}
\left \{
{\mathcal N}(\bm{\phi}_i \mid \bm{0}, \hat{R}) 
P(\sigma_i(0))\prod_{t=0}^{T-1} 
\left (
\delta(\sigma_i(t+1)-\sgn(h_i(t))) e^{i\ell_i(t)\sigma_i(t)} \right )\right . \cr
&\quad \left . \times 
 \exp\left (i \sum_{t=0}^{T-1} 
 g_i(t) \left (h_i(t) - 
 \frac{\xi_i^0 m(t)^{p-1}}{(p-1)!} -\phi_i(t) 
 + \sum_{s} \hat{S}(s,t) \sigma_i(s)\right ) 
 + \frac{1}{2} \bm{\sigma}_i \hat{Q} \bm{\sigma}_i \right )
\right \}
\label{eq:Omega} \\
&\propto \mathop{\rm tr}_{\bm{h}_i, \bm{\sigma}_i, \bm{\phi}_i}
\left \{
{\mathcal N}(\bm{\phi}_i \mid \bm{0}, \hat{R}) 
P(\sigma_i(0))\prod_{t=0}^{T-1} 
\left (
\delta(\sigma_i(t+1)-\sgn(h_i(t))) e^{i\ell_i(t)\sigma_i(t)} \right )\right . \cr
&\quad \left . \times 
 \prod_{t=0}^{T-1} 
 \delta  \left (h_i(t) - 
 \frac{\xi_i^0 m(t)^{p-1}}{(p-1)!} -\phi_i(t) 
 + \sum_{s} \hat{S}(s,t) \sigma_i(s)\right )
\times  \exp \left ( 
\frac{1}{2} \bm{\sigma}_i \hat{Q} \bm{\sigma}_i \right )
\right \}. \nonumber 
\end{align}
We employed a Gaussian integral formula
\begin{align}
\exp\left (\frac{1}{2}(i\bm{g}_i)^\top \hat{R} (i\bm{g}_i) \right )
= \int {\mathcal N}(\bm{\phi}_i \mid \bm{0}, \hat{R}) \exp \left (-\sum_{t=0}^{T-1} ig_i(t)\phi_i(t) \right )
{\mathrm d} \bm{\phi}_i
\end{align}
to have the expression of Eq. \eqref{eq:Omega}.

\subsection{Extremization condition}

Extremizing the right-hand side of Eq.~\eqref{eq:free_energy} in the limit 
$\tilde{\bm{\ell}} \to 0$ yields
\begin{align}
\hat{Q}(t,s) &= \alpha \left(
\frac{Q(t,s)^{p-2} R(t,s)}{(p-2)!}
+ \frac{Q(t,s)^{p-3}\big(S(t,t)S(s,s)+ S(t,s)S(s,t)\big)}{(p-3)!}
\right), \label{eq:Q_hat_saddle} \\
\hat{R}(t,s) &= \frac{\alpha Q(t,s)^{p-1}}{(p-1)!}, \\
\hat{S}(t,s) &= \frac{\alpha Q(t,s)^{p-2}(1 + \delta(t,s)) S(s,t)}{(p-2)!}, \\
Q(t,s) &= \frac{1}{N}\sum_{i=1}^N \mathbb{E}^{\Omega_0} [\sigma_i(t)\sigma_i(s)], \\
R(t,s) &= \frac{1}{N}\sum_{i=1}^N \mathbb{E}^{\Omega_0} [(ig_i(t))(ig_i(s))], \label{eq:R_saddle} \\
S(t,s) &= \frac{1}{N} \sum_{i=1}^N \mathbb{E}^{\Omega_0} [\sigma_i(t) (ig_i(s))], \label{eq:S_saddle}
\end{align}
where $\mathbb{E}^{\Omega_0}[\cdots]$ denotes the expectation with respect to 
the measure induced by the integrand of Eq.~\eqref{eq:Omega} in the limit 
$\tilde{\bm{\ell}} \to 0$.

Several remarks are in order:
\begin{itemize}

\item 
The normalization property
\[
\lim_{\tilde{\bm{\ell}}\to \bm{0}} 
\mathbb{E}_{\bm{J}}\big[Z(\tilde{\bm{\ell}} \mid \bm{J})\big] = 1
\]
holds even in the presence of any time-dependent external fields $h_i^{\rm ex}(t)$ added to Eq.~\eqref{eq:field}. 
Furthermore, Eq.~\eqref{eq:Omega} implies
\begin{align}
\mathbb{E}^{\Omega_0}[(ig_i(t))(ig_i(s))]
=
\lim_{\tilde{\bm{\ell}}\to \bm{0},\, \bm{h}^{\rm ex}\to \bm{0}}
\frac{\partial^2}{\partial h_i^{\rm ex}(t)\, \partial h_i^{\rm ex}(s)}
\mathbb{E}_{\bm{J}}\big[Z(\tilde{\bm{\ell}} \mid \bm{J})\big].
\end{align}
Together with Eq.~\eqref{eq:R_saddle}, this leads to
\begin{align}
R(t,s) = 0
\label{eq:R_zero}
\end{align}
for all $t$ and $s$.

\item 
Equation~\eqref{eq:Omega} implies
\begin{align}
\mathbb{E}^{\Omega_0}[\sigma_i(t)(ig_i(s))]
&= -\mathbb{E}^{\Omega_0} \left[
\frac{\partial \sigma_i(t)}{\partial \phi_i(s)}
\right] \nonumber \\
&=
\begin{cases}
-\displaystyle\sum_{u=0}^{t-1} 
\mathbb{E}^{\Omega_0} [\sigma_i(t)\phi_i(u)]\, \hat{R}^{-1}(u,s),
& s < t, \\
0, & s \ge t,
\end{cases}
\label{eq:sigma_phi}
\end{align}
where we used causality, i.e., $\partial \sigma_i(t)/\partial \phi_i(s)=0$ for $s \ge t$, 
together with Stein's lemma.

For $(t,s)=(1,0)$, this can be evaluated analytically as
\begin{align}
\mathbb{E}^{\Omega_0}\left[
\frac{\partial \sigma_i(1)}{\partial \phi_i(0)}
\right]
&=
\int \mathcal{N}\!\left(\phi_i(0)\mid 0, \frac{\alpha}{(p-1)!}\right)
\frac{\partial}{\partial \phi_i(0)}
\sgn\!\left(
\frac{\xi^0 m(0)^{p-1}}{(p-1)!} + \phi_i(0)
\right)
\, d\phi_i(0) \nonumber \\
&=
2 \int \mathcal{N}\!\left(\phi_i(0)\mid 0, \frac{\alpha}{(p-1)!}\right)
\delta\!\left(
\frac{\xi^0 m(0)^{p-1}}{(p-1)!} + \phi_i(0)
\right)
\, d\phi_i(0) \nonumber \\
&=
\sqrt{\frac{2(p-1)!}{\pi \alpha}}
\exp\left(
-\frac{m(0)^{2(p-1)}}{2(p-1)!}
\right),
\end{align}
which yields Eq.~\eqref{eq:S10}.

\item 
Substituting Eq.~\eqref{eq:sigma_phi}, together with 
Eqs.~\eqref{eq:S_saddle} and \eqref{eq:R_zero}, into 
Eq.~\eqref{eq:Q_hat_saddle}, we obtain
\begin{align}
\hat{Q}(t,s) = 0
\end{align}
for all $t$ and $s$.

\end{itemize}

Finally, evaluating $\mathbb{E}^{\Omega_0}[\cdots]$ via Monte Carlo sampling with a large number of realizations 
leads to the macroscopic dynamics presented in Sec.~\ref{subsec:single_site}.

\bibliographystyle{apsrev4-2}
\bibliography{pHopfieldNotes}

\end{document}